\pdfoutput=1
\documentclass[11pt]{article}

\usepackage[preprint]{acl}

\usepackage{times}
\usepackage{latexsym}

\usepackage[T1]{fontenc}
\usepackage[utf8]{inputenc}

\usepackage{microtype}

\usepackage{inconsolata}

\usepackage{graphicx}
\usepackage{amsmath}
\usepackage{amssymb}
\usepackage{amsfonts}
\usepackage{algorithm}
\usepackage{algpseudocode}
\usepackage{booktabs}
\usepackage{verbatim}
\usepackage{multirow}
\usepackage{enumitem}
\usepackage{tcolorbox}
\usepackage{float}
\usepackage{ulem}
\usepackage{soul}
\usepackage{color}
\usepackage{tikz}
\title{Generating Querying Code from Text for \\ 
Multi-Modal Electronic Health Record}

\author{Mengliang Zhang \\
  The University of Texas at Arlington, CSE Department\\
  }

\begin{document}

\maketitle
\begin{abstract}
Electronic health records (EHR) contain extensive structured and unstructured data, including tabular information and free-text clinical notes. Querying relevant patient information often requires complex database operations, increasing the workload for clinicians. However, complex table relationships and professional terminology in EHRs limit the query accuracy. In this work, we construct a publicly available dataset, TQGen, that integrates both \textbf{T}ables and clinical \textbf{T}ext for natural language-to-query \textbf{Gen}eration. To address the challenges posed by complex medical terminology and diverse types of questions in EHRs, we propose TQGen-EHRQuery, a framework comprising a medical knowledge module and a questions template matching module. For processing medical text, we introduced the concept of a toolset, which encapsulates the text processing module as a callable tool, thereby improving processing efficiency and flexibility. We conducted extensive experiments to assess the effectiveness of our dataset and workflow, demonstrating their potential to enhance information querying in EHR systems.
\end{abstract}

%However, existing approaches often fail to account for the complexities of EHR data, such as intricate table relationships, lengthy clinical narratives, and the extensive use of specialized medical terminology.

\section{Introduction}

Electronic Health Records (EHR)~\cite{johnson2016mimiciii, pollard2018eicu, johnson2023mimiciv} contain a vast amount of tabular and textual information about patients. Retrieving this information often requires complex database queries, posing a challenge for clinicians without specialized database expertise. Converting natural language queries into structured database queries can significantly enhance the efficiency of medical professionals. Previous research has explored table-based text-to-SQL models, including sequence-to-sequence approaches~\cite{dong2016language} and large language model-based methods. With the emergence of large-scale EHR datasets, several works~\cite{raghavan2021emrkbqa,lee2022ehrsql} have introduced table-based text-to-SQL datasets tailored for EHRs. While these methods have demonstrated promising performance, they still have certain limitations.

\begin{figure}[htbp]  % htbp 控制图片浮动的位置
    \centering  % 图片居中
    \includegraphics[width=0.4\textwidth]{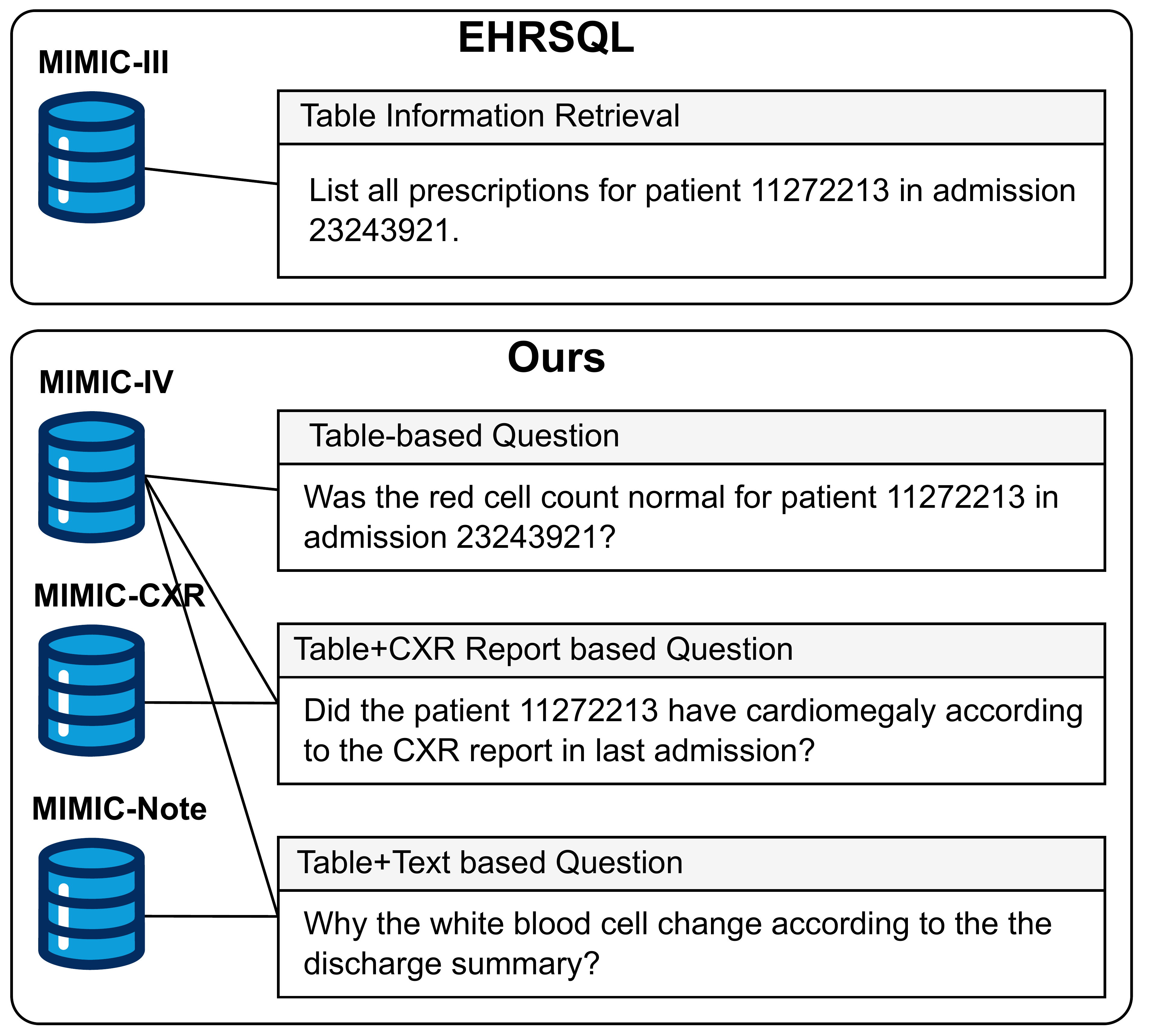}  % 调整图片大小并插入图片
    \caption{Previous work (such as EHRSQL) focuses only on tabular information, we introduce textual data within tables and leverage multi-modal interactions to create queries.}  % 图片标题
    \label{fig:introduction}  % 用于引用图片
\end{figure}

%%Electronic health records (EHR)~\cite{johnson2016mimiciii, pollard2018eicu, johnson2023mimiciv} contain a large amount of tabular and textual information about patients. When retrieving this information, doctors need to perform specialized database queries, which adds to the workload of healthcare professionals. Some previous studies have aimed to design models that convert natural language input into query statements, enabling database table queries, and thereby lowering the barrier to database operations.

EHR contains both structured tabular data and unstructured textual information, such as radiology reports and discharge summaries. These texts may be stored directly within table columns as long-form narratives or referenced via external links. Previous studies~\cite{lee2022ehrsql} have primarily focused on querying structured tables without integrating text comprehension, see Fig~\ref{fig:introduction} while others~\cite{kweon2024ehrnoteqa} have explored long-text processing but lacked the capability to handle multimodal table-text queries in EHRs. Additionally, some works~\cite{bae2023ehrxqa} have proposed VQA tasks based on tables and chest X-ray (CXR) images. However, in clinical practice, radiology reports are already generated by radiologists, making direct image-based queries less practical. To bridge these gaps, we construct a comprehensive table-text query generation dataset by integrating structured data from MIMIC-IV~\cite{johnson2023mimiciv}, radiology reports from MIMIC-CXR~\cite{johnson2019mimiccxr}, and discharge summaries from MIMIC-Note~\cite{johnson2023mimic}, facilitating more effective and clinically relevant EHR retrieval.

Besides, EHR contain numerous specialized medical terms, often represented inconsistently (e.g., ‘red blood cell’ vs. ‘RBC’). This variation complicates converting natural language queries into accurate database queries. To address this, we incorporate medical knowledge to identify and map specialized terms in clinician queries to corresponding database terms. Furthermore, the combination of tabular data and text within tables poses challenges for query code generation. To address this, we introduce the concept of a toolset, encapsulating medical text processing functions into callable tools. When the model detects the need to interpret textual data such as CXR reports, it invokes these tools, thereby extending the modality coverage of text-to-SQL systems.

Moreover, the current query code generation for EHR data lacks a standardized processing framework. In response, we propose a query code processing framework, TQGen-EHRQuery, which leverages large models as the foundational component. This framework encompasses several key modules, including table content description, medical terminology matching, question template matching, query generation prompts, and query execution validation. These introductions are described in Section~\ref{sec:methodology}.

%Converting natural language into query statements must ensure query stability across different healthcare professionals and language styles. Therefore, it is crucial for the model to produce consistent query results when faced with natural language inputs that have the same meaning but different expressions.

%Some methods generate SQL queries multiple times and select the most frequently occurring result through voting. Others utilize multiple LLMs or agents to generate SQL queries and vote for the optimal answer. In constructing our dataset, we generated multiple variations of the same question template to simulate different expression styles. Additionally, we experimented with generating SQL queries multiple times and selecting the most frequent result through voting.

Our contributions are as follows.

\begin{enumerate}[itemsep=0pt, parsep=0pt, topsep=0pt]
    \item We have constructed a natural language query dataset that integrates tabular electronic health record (EHR) data with medical text records. This data set expands the textual modality, making natural language queries for EHRs more aligned with real-world scenarios.
    \item We propose an EHR query processing framework based on a large language model, incorporating a medical knowledge module, question template matching, and other components to enhance query accuracy. Notably, we introduce the toolset concept and design text processing tools to extend query modality.
    \item We evaluated our workflow on the proposed dataset, demonstrating the effectiveness of our approach.
\end{enumerate}

The remainder of the paper is organized into several sections. Section~\ref{sec:related work} discusses existing related work, Section~\ref{sec:dataset construction} describes the TQGen dataset generation, Section~\ref{sec:methodology} presents the framework for EHR multi-modal query generation, Section~\ref{sec:experiment} presents the experiments and results, Section~\ref{sec:limitation} discusses the limitation of our work, Section~\ref{sec:conclusion} concludes the work and discusses possible future work.

\section{Related Work}
\label{sec:related work}

Classic benchmark datasets such as WikiSQL~\cite{zhong2017seqsql} and Spider 1.0~\cite{yu2018spider} have significantly contributed to text-to-SQL task development. However, with the rise of large language models (LLMs), these datasets have shown limitations, such as lacking domain-specific knowledge and large-scale table structures. Recent benchmarks like DB-GPT-Hub~\cite{zhou2024dbgpthub} and BIRD~\cite{li2024can} address real-world challenges, including domain-specific knowledge, large-scale tables, and data noise, offering new directions for text-to-SQL research.

In the EHR domain, text-to-SQL tasks focus on extracting information from medical record tables by translating natural language into SQL or other query languages. Wang et al.\cite{wang2020texttosql} introduced TREQS, which performs text-to-SQL tasks on MIMIC-III. Pampari et al.\cite{raghavan2021emrkbqa} developed emrKBQA, a large-scale text-to-logical-form dataset for patient-specific QA on MIMIC-III. Lee et al.\cite{lee2022ehrsql} presented EHRSQL, a text-to-SQL dataset based on MIMIC-III and eICU\cite{pollard2018eicu}, incorporating time-sensitive and unanswerable queries. EHRNoteQA~\cite{kweon2024ehrnoteqa} provides QA tasks from discharge summaries, serving as a long-text benchmark based on MIMIC-IV data.

\begin{figure*}[htbp]  % htbp 控制图片浮动的位置
  \centering  % 图片居中
  \includegraphics[width=0.9\textwidth]{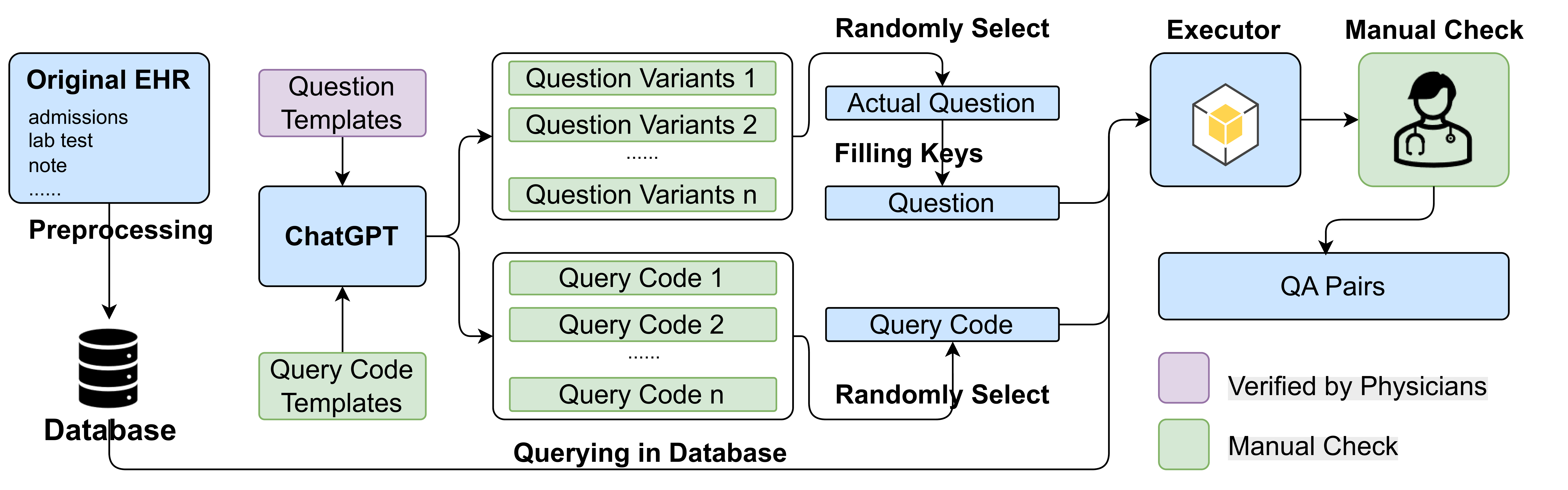}  % 调整图片大小并插入图片
  \caption{The pipeline of dataset construction. After preprocessing to the EHR data, we create template for question and query code, then execute the code to obatin the question-answer (QA) pairs. The purple boxes indicate that the content has been verified by physicians, and green boxes indicate that it has been manually checked.}  % 图片标题
  \label{fig:dataset construction}  % 用于引用图片
\end{figure*}

\section{Dataset Construction}
\label{sec:dataset construction}

This study utilizes MIMIC-IV, integrating radiology reports from MIMIC-CXR~\cite{johnson2019mimiccxr} and discharge summaries from MIMIC-Note~\cite{johnson2023mimic}. These reports are embedded as text or hyperlinks within structured tables, facilitating data association and analysis. A detailed dataset description is provided in Appendix~\ref{sec:ehr dataset introduction}.

%\textcolor{red}{This study utilizes MIMIC-IV, integrating radiology reports from MIMIC-CXR~\cite{johnson2019mimiccxr} and discharge summaries from MIMIC-Note~\cite{johnson2023mimic}. These reports are embedded as text or hyperlinks within structured tables, facilitating data association and analysis. A detailed dataset description is provided in Appendix~\ref{sec:ehr dataset introduction}.}

%MIMIC-IV (v2.2) ~\cite{johnson2023mimiciv} dataset is a large, freely accessible relational database of de-identified health-related data (e.g., diagnoses, procedures, and treatments) associated with 50,920 patients who stayed in critical care units of Beth Israel Deaconess Medical Center (BIDMC) between 2008-2019.

%MIMIC-CXR ~\cite{johnson2019mimiccxr} is a large-scale publicly available dataset of 377,110 chest radiographs associated with 227,827 imaging studies sourced from the BIDMC between 2011-2016. MIMIC-CXR can be linked to MIMIC-IV using lookup tables that connect patient identifiers.

%MIMIC-IV-Note ~\cite{johnson2023mimiciv} is a deidentified free-text clinical note dataset linked to the MIMIC-IV database. It includes 331794 discharge summaries from 145915 patients (hospital \& emergency department), 2321355 radiology reports from 237427 patients. All notes are deidentified per HIPAA Safe Harbor standards.

\subsection{Preprocessing}
\label{sec:preprocessing}

%%% original
In this study, we integrated multi-source datasets by standardizing and linking the raw data using the unique patient identifier (subject\_id) and hospital admission ID (hadm\_id). Additionally, intensive care unit stay ID (stay\_id) and chest X-ray study ID (study\_id) were utilized to identify eligible patient cohorts, ensuring a high-quality data foundation for subsequent analyses.

%%% original
During preprocessing, all table fields underwent type validation and standardization. Text data was converted to lowercase for consistency. To improve query efficiency, related tables were merged (e.g., integrating diagnoses with d\_icd\_diagnoses). Detailed preprocessing methods are provided in Appendix~\ref{sec:appendix dataset construction}.

%The CXR table was streamlined to retain key diagnostic information, such as imaging angles and disease labels, while removing non-essential details.

%In this section, we describe the process of building the dataset. We integrate EHR tables, CXR images, and EHR discharge reports based on patient information. Then, we provide a detailed explanation of the creation of question templates and the corresponding query statements. Finally, we discuss dataset generation and validation results.

%To cross-reference EHR tables, CXR radiology reports, and discharge summaries, we need to establish an index across different data sources based on patient information. For this purpose, we use "subject\_id" and "hadm\_id" to link different datasets, where "subject\_id" represents the patient ID and "hadm\_id" represents the admission ID. This approach allows us to retrieve the radiology reports corresponding to a patient’s CXR images, along with other tabular data. Details of the data processing steps can be found in Appendix~\ref{sec:appendix}.

\begin{table*}
  \centering
  \small
  \begin{tabular}{lcccc}
    \hline
    \textbf{Dataset}& \textbf{Table} & \textbf{Text} & \textbf{\# of Tables} & \textbf{\# of Questions}\\
    \hline
    TREQS~\cite{wang2020texttosql}          & $\checkmark$ & $\times$                   &  5 & 10k            \\
    EHRSQL~\cite{lee2022ehrsql}       & $\checkmark$  & $\times$                   & 13.5  & 24k              \\
    EHRXQA~\cite{bae2023ehrxqa}        & $\checkmark$ & $\times$           & 18 & 46k           \\
    EHRAgent~\cite{shi2024ehragent}  & $\checkmark$ & $\times$      &  10     &  2k         \\
    EHRNote~\cite{kweon2024ehrnoteqa}        & $\times$ & $\checkmark$           & 1 & 1k           \\
    TQGen(Ours)        & $\checkmark$ & $\checkmark$       & 18  & 12k              \\
    \hline
  \end{tabular}
  \caption{Dataset comparison with other EHR-based text-to-query dataset.}
  \label{tab:dataset compare} 
\end{table*}

\subsection{Question Template}
\label{sec:question template}

%%% original
%{\color{red}\sout{Since the data involves multi-modal information from different modalities (tables, text), we define the scope of question templates using two dimensions: table-based and text-based question.}}

%%% original
%{\color{red}\sout{Table-based questions are associated with structured information from EHR tables. These questions address patient demographics, diagnoses, procedures, medications, and other clinical details typically recorded in a structured EHR format. The dataset offers a rich collection of questions derived from EHR tables, making it a highly valuable resource in this context. We utilized question templates from the MIMIC-III version of EHRSQL, adapting them to align with the MIMIC-IV schema with necessary modifications. Approximately 100 question templates were constructed for table-based queries, with examples provided in Appendix~\ref{sec:appendix dataset construction}.}}

%%% original
%The text-based question involves questions derived from discharge summaries and CXR radiology reports. These questions cover patient conditions, changes in the CXR radiology reports, the patient’s admission history, discharge diagnoses, and more. We enhanced the templates to handle queries specific to individual patients and for comparisons between two consecutive CXR studies. Approximately 80 question templates were constructed for text-based queries, with examples provided in Appendix~\ref{sec:appendix dataset construction}.

We first developed a series of question templates related to EHR information, covering topics such as patient details, laboratory indicators, medication usage, length of hospitalization, discharge information, and radiology reports. In consultation with physicians, we refined and optimized these templates to ensure their clinical relevance. Subsequently, for each template, we used ChatGPT~\cite{achiam2023gpt} to generate multiple alternative phrasings, which were manually reviewed and selected, resulting in approximately ten distinct variants per template. This approach enhances the diversity and robustness of the question formulations. For table-based questions, we referenced the question types in EHRSQL~\cite{lee2022ehrsql} and, based on physicians’ advice, added additional questions related to laboratory parameters, such as their values, changes, statistical parameters, and whether they were within normal ranges. These questions were considered to have important clinical value.

For text-based questions, we first analyzed the types of information contained in the discharge summary, such as the patient’s medical history, reason for admission, discharge prescriptions, and discharge status. Based on this analysis, we designed corresponding question templates, ensuring their clinical meaning with the guidance of physicians. Additionally, we observed that discharge summaries often mention laboratory indicators and medication usage, which partially overlap with the data available in the tables. Kwon et al. \cite{kwon2024ehrcon} also highlighted the disagreement between EHR table data and textual information. However, the laboratory indicators presented in the text are incomplete, with some date information omitted for privacy reasons. Therefore, to obtain more comprehensive data on laboratory indicators and medication usage, we designed prompts that direct the model to retrieve answers from the tables. For text-based questions, we explicitly instructed the model to extract answers from the textual data within the question templates, thereby minimizing potential biases caused by inconsistencies between data sources.

After multiple rounds of discussions and revisions with two physicians, we developed approximately 100 question templates related to table data and about 80 templates related to text data. For each template, the actual question was randomly selected from its variants and populated with the relevant information based on the specific template, ensuring both diversity and clinical relevance in the generated questions. Specific examples of these question templates can be found in Appendix~\ref{sec:appendix dataset construction}.

When generating actual questions, we first select a question template and randomly choose one of its variants as the final question. We then fill the keys with specific values, such as subject\_id or medication name, ensuring that it contains real data.

\subsection{Query Answer Generation}
\label{sec:answer generation}

%%% original
%%{\color{red}\sout{In Fig~\ref{fig:dataset construction}, for each question template, we design a corresponding query statement, which is executed after inserting relevant keywords to query results. For questions requiring answer extraction from long-text data, we input both the text and the corresponding question into a large pre-trained model, followed by manual verification of the generated responses. In cases where the query yields no valid information (e.g., inquiries about examinations not performed on a given patient), predefined prompts are used as response outputs.}}

In Fig~\ref{fig:dataset construction}, for each question template, we manually designed the corresponding query code and used ChatGPT~\cite{achiam2023gpt} to generate multiple variations of the queries to ensure diversity. These generated query codes were then manually reviewed to ensure their syntactical correctness, followed by execution in the database to verify the accuracy of the results. In addition, we randomly selected 500 query codes, executed them and returned the results for further evaluation of their correctness. For table-based questions, we relied on manually written queries and validated their reliability through query execution. For text-based questions, we utilized the offline-deployed Qwen2.5 70B~\cite{yang2024qwen} model to process discharge summaries and extract answers relevant to the queries.

To validate the generated answer, we randomly selected 50 discharge summaries, each paired with ten different questions, and invited two physicians to review the responses. Specific examples of these questions are provided in the appendix. In practice, some questions may not have an answer—for example, querying a patient’s red blood cell count when no such test has been performed. Prior studies~\cite{lee2024overview} have also highlighted the issue of unanswerable questions. To address this, we predefined the response text as “No corresponding information found,” ensuring consistent handling of missing data. This approach enhances model robustness by preventing the generation of incorrect or unreliable answers, thereby improving the overall reliability and practicality of the question-answering system.

\begin{figure*}[htbp]  % htbp 控制图片浮动的位置
  \centering  % 图片居中
  \includegraphics[width=0.9\textwidth]{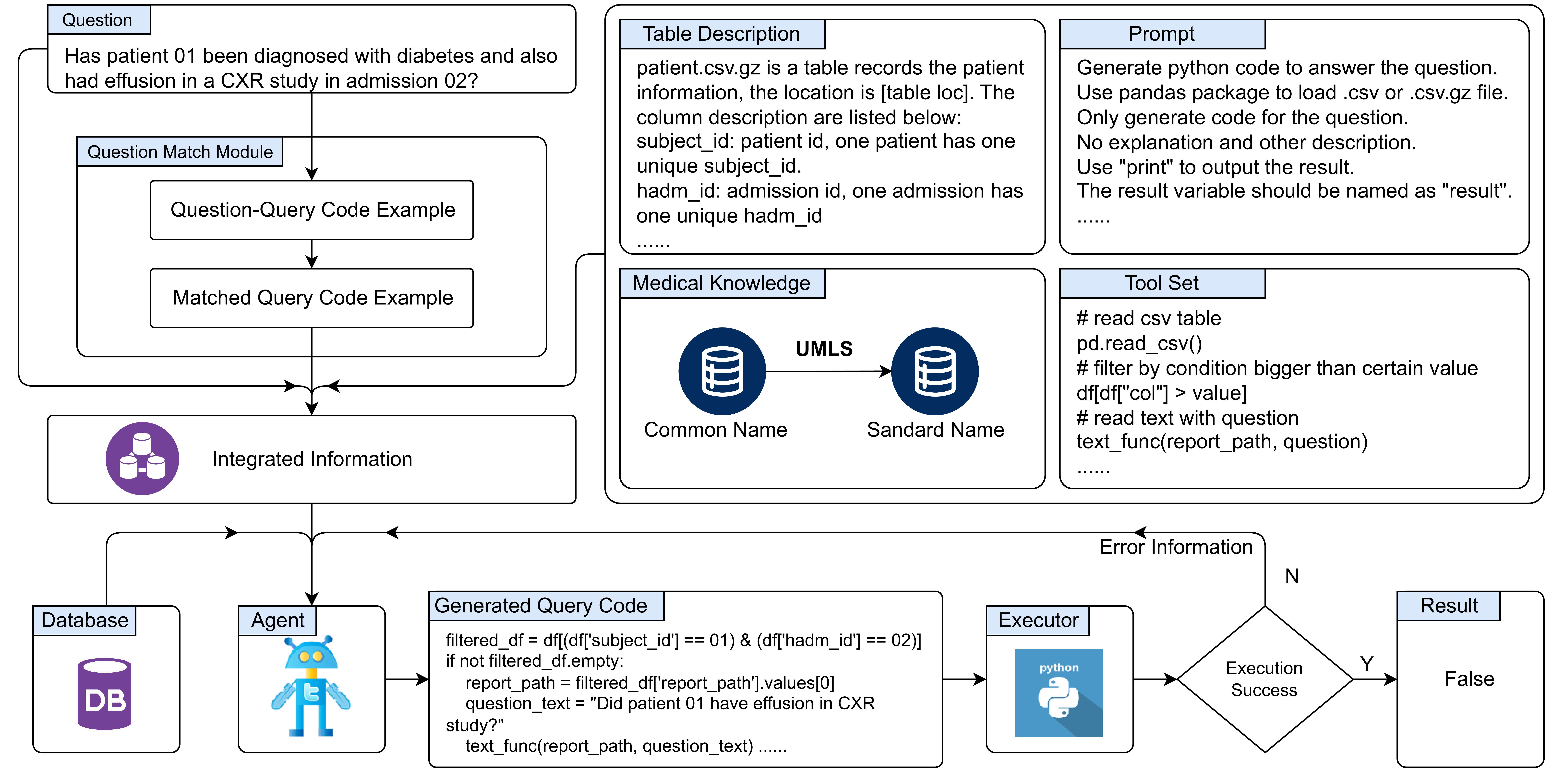}  % 调整图片大小并插入图片
  \caption{The framework of generating query code from question text. We use python code as example.}  % 图片标题
  \label{fig:framework}  % 用于引用图片
\end{figure*}

\begin{figure}[htbp]  % htbp 控制图片浮动的位置
  \centering  % 图片居中
  \includegraphics[width=0.4\textwidth]{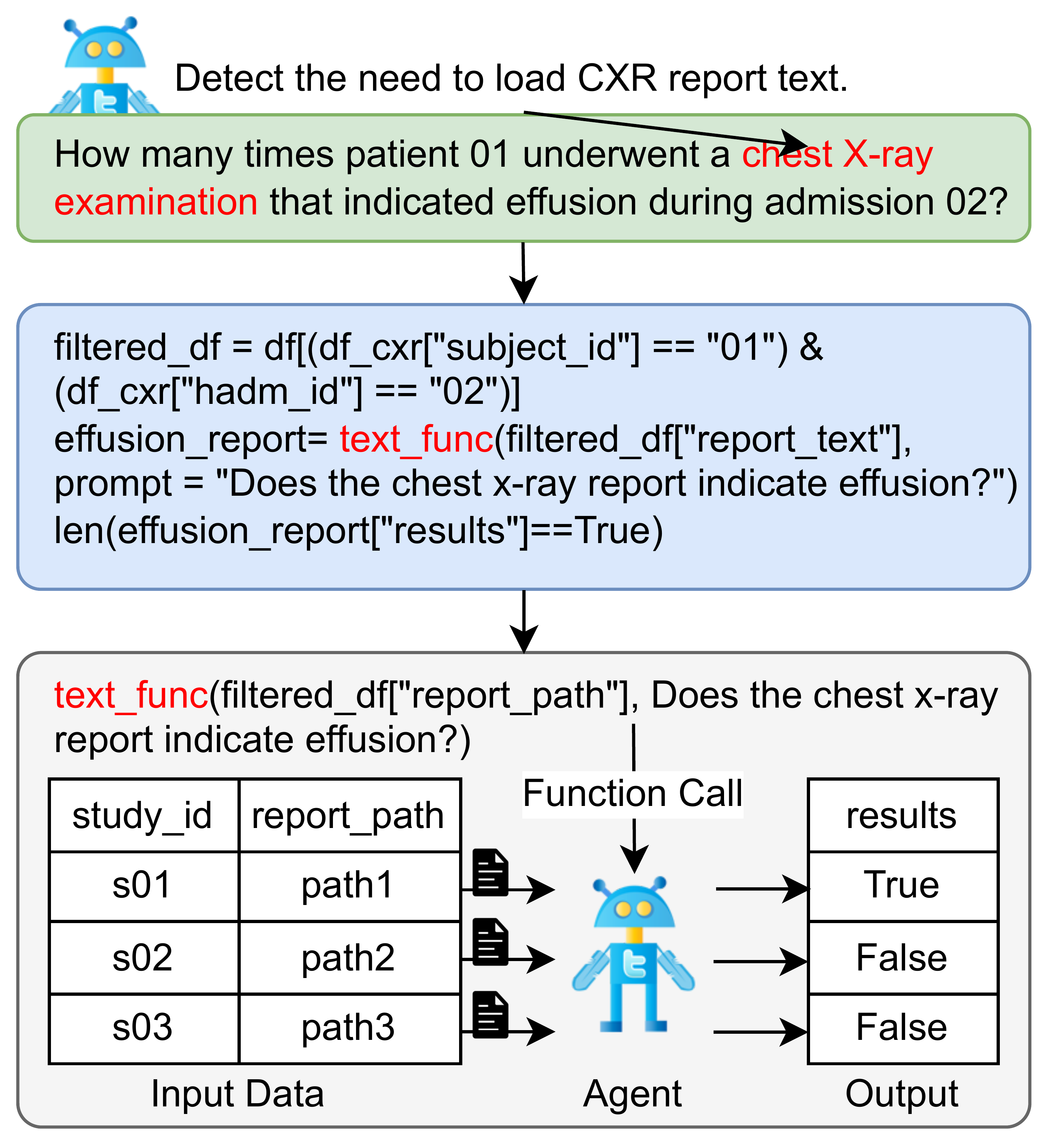}  % 调整图片大小并插入图片
  \caption{The method to call function to process text.}  % 图片标题
  \label{fig:text_func}  % 用于引用图片
\end{figure}

\subsection{Dataset Distribution}

We conduct a comparative analysis of previous datasets, as presented in Table~\ref{tab:dataset compare}, and provide statistics on the distribution of question counts across different modalities, as shown in Table~\ref{tab:dataset split}. Furthermore, we perform a classification analysis based on question complexity, categorizing questions into two levels: Level I for questions with no more than three constraint conditions (eg. subject\_id, diagnoses\_name), and Level II for those with more than three.

\section{Methodology}
\label{sec:methodology}

\subsection{Preliminary}
\label{sec:preliminary}

In this work, we focus on addressing health-related queries using information from structured EHRs. The reference EHR, denoted as $\mathcal{D} = \{D_0, D_1, ...\}$, $D_i$ represents the $i_{th}$ table in database, and $\mathcal{C}^{i} = \{C_0^{i}, C_1^{i}, ...\}$ corresponds to the column description with in $D_{i}$. Given an EHR-based clinical question $q \in Q$, the objective is to extract the final answer by utilizing the information with both $\mathcal{D}$ and $\mathcal{C}$. The equation is:

%%{\color{red}\sout{We further develop the planning process of LLM as an autonomous agent in EHR question answering. For initialization, the LLM agent is equipped with a set of pre-built tools $\mathcal{T} = \{T_0, T_1, ... \}$ to interate with the EHR database $\mathcal{D}$. For example, the "SUM", "COUNT" functions can be regarded as tools in SQL {\color{red}\sout{querylanguage}} \textcolor{blue}{query language}. The query code generation can be regarded as a combination of tools. Thus we generate the sequence by the following policy: $p_{q} \sim p(f_{1}, ... f_{t} | q, \mathcal{D}, \mathcal{C}, \mathcal{T})$, where $q \in \mathcal{Q}, f_{t} \in {\mathcal{T}}$. The final output is obtained by executing the function sequence:}}

\vspace{-5mm}
\begin{equation}
    y \sim \text{EXECUTOR}(q, f_{1}, ... f_{t}, \mathcal{D}, \mathcal{C})
\end{equation}

where the $\text{EXECUTOR}$ is the query code executor interacting with EHR database.

%%{\color{red}\sout{We then trace the outcome of each interaction back to the LLM agent, which can be either a successful execution result or an error message, to iteratively refine the generated code-based plan. This interactive process is a multi-turn conversation between the planner and executor, which leverages the high-level reasoning capabilities of the LLM to optimize plan refinement and execution.}}

%How to use footnote.
%Footnotes are inserted with the \verb|\footnote| command.\footnote{This is a footnote.}

\subsection{Modules}

Building upon the dataset introduced in Section~\ref{sec:dataset construction} and the fundamental principles of EHR querying, we propose a framework named TQGen-EHRQuery. This framework is specifically optimized for EHR and is designed to efficiently translate natural language into query code. TQGen-EHRQuery consists of several key modules, including the table description module, matching module, tool set module, and code inspection module, ensuring the accuracy and reliability of query generation.

\noindent \textbf{Table Description.} In the process of converting natural language into query code, table description is a key module responsible for establishing connections between natural language queries and the structured schema of relational databases. It helps the model accurately map query terms to database columns, tables, or values, thereby improving the generation of query code. We present an example of table description in Appendix~\ref{sec:prompt detail}.

\noindent \textbf{Matching Module.} When parsing a question, the model matches text with values in the table. For example, in “What is the highest red blood count value of patient 01 in admission 02?”, table description helps the model locate the subject\_id, hadm\_id, and label columns in the labtest table to query the patient’s value and select the highest one. However, medical terms like red blood count may appear as RBC or red blood cell in different inputs, complicating the mapping process. To address this, we established a standard terminology library to ensure consistent mapping of input terms. Leveraging UMLS ~\cite{bodenreider2004unified} standardized medical terms, rbc and red blood cell are uniformly mapped to red blood cell.

To generate queries code, we retrieve historical query templates based on semantic similarity. Pre-stored queries and their corresponding questions are collected and stored. We calculate the similarity between the input question and existing ones using a pre-trained BERT~\cite{reimers2019sentencebert} model. The most similar questions are then retrieved, and their corresponding query templates are extracted. For large template databases, Faiss~\cite{douze2024faiss} is used for efficient similarity search. If no exact SQL template is found, multiple similar templates are combined to automatically generate the SQL query structure.

\noindent \textbf{Tool Set Module.} Since some tables embed links to long texts or directly embed long texts, and the query statement cannot directly extract and understand the corresponding content from the long text, we designed a text understanding tool. When the model parses the input question and finds that the query content involves long texts such as radiology reports or discharge reports, we use the text understanding tool. This tool is packaged into a function. Its input is the long text and the question, and the output is the corresponding value.

In this work, we propose an automatic method for generating dynamic prompts for a text understanding function, $\text{Text\_Func}$, based on the original query. For a given question, we extract key entities such as patient ID, admission ID, and medical conditions using a table description module. For example, from the query “Count the number of times that patient 01 had a CXR check indicating effusion in admission 02”, we extract patient\_id = 01, admission\_id = 02, and condition = effusion. Using this extracted information, we dynamically generate a prompt to guide $\text{Text\_Func}$ in retrieving relevant data from medical records. For instance, the prompt would be: “Does the chest x-ray report of patient 01 in admission 02 indicate effusion?”. Figure~\ref{fig:text_func} illustrates the pipeline for using $\text{Text\_Func}$ to process the text.

%\textbf{Query Code Executor.} 
\begin{algorithm}
\footnotesize  % 或 \small
\caption{Algorithm Framework}
\label{alg:medical_process}
\begin{algorithmic}
    \State \textbf{Input:} EHR database $\mathcal{D}$, question $q \in \mathcal{Q}$, column descriptions $\mathcal{C}$, tools $\mathcal{T}$, samples $\mathcal{Q}_{s}$, knowledge $\mathcal{M}$, prompt $\mathcal{P}$.
    \State Guided prompt: $\mathcal{I} = [\mathcal{C}, \mathcal{M}, \mathcal{P}]$
    \State Initialize: $k = 1$, $\text{flag} = 0$
    
    \State \textcolor{blue}{\% Match similar question examples}
    \State $q_{sim} = \text{TopK}_{\max}(q, \mathcal{Q}_s)$
    
    \State \textcolor{blue}{\% Generate Query Code}
    \State $S(q) = \text{LLM}(\mathcal{I}, \mathcal{T}, q, q_{sim})$
    
    \While{$k \leq K$ \textbf{and} $\text{flag} = 0$}
        \State \textcolor{blue}{\% Code Execution}
        \State $O(q) = \text{EXECUTOR}(S(q))$
        
        \If{$O(q)$ contains error}
            \State $S(q) = \text{LLM}(S(q), \mathcal{I}, \mathcal{T}, q, \text{error})$
            \State $k = k + 1$
        \Else
            \State $\text{flag} = 1$
        \EndIf
    \EndWhile
    
    \State \textbf{Output:} $O(q)$ (final answer or output)
\end{algorithmic}
\end{algorithm}

\noindent \textbf{Code Inspection Module.} The code executor automatically extracts the code from the LLM agent output and
executes it within the local environment: 

\vspace{-5mm}
\begin{equation}
    O(q) = \text{EXECUTOR}(S(q)))
\end{equation}

After execution, it sends the results of execution back to the LLM agent for potential plan refinement and further processing.

We observe that the generated query statements do not always execute successfully. To address this issue, we incorporate a repair module to refine queries that fail during execution. When the generated query statement $S(q)$ encounters an error in the executor, we identify potential issues such as incorrect file path references, column mismatches, or erroneous value assignments. To improve query accuracy, we collect the error messages returned by the executor along with the original question, the generated query, guiding prompts, and relevant toolbox resources. This information is then fed back into the LLM agent iteratively until a valid query is produced or the predefined query attempt limit is reached. The equation is as follows:

\vspace{-5mm}
\begin{equation}
    S(q) = \text{LLM}(S(q), \mathcal{I}, \mathcal{T}, q, \text{error\_info})
\end{equation}

The logic of the code inspection module can be found in Algorithm \ref{alg:medical_process}.

\begin{table*}[t]
    \centering
    \small
    \scalebox{0.9}{  % adjust table 90%
        \begin{tabular}{lccccccc}
            \toprule
            & & \multicolumn{2}{c}{EM} & \multicolumn{2}{c}{EX} & \multicolumn{2}{c}{LLM-based score} \\
            \cmidrule(r){3-4} \cmidrule(r){5-6} \cmidrule(r){7-8}
            Model & Size & Level I & Level II & Level I & Level II & Level I & Level II \\
            \midrule
            \multirow{3}{*}{Qwen 2.5~\cite{yang2024qwen}} & 7b & 0.53 & 0.48 & 0.80 & 0.63 & 7.85 & 6.04 \\
            & 14b & 0.58 & 0.53 & 0.82 & 0.75 & 8.53 & 6.57 \\
            & 32b & 0.61 & 0.59 & 0.89 & 0.80 & 9.05 & 7.12 \\
            \midrule
            \multirow{3}{*}{Qwen 2.5 Code~\cite{hui2024qwen2}} & 7b & 0.56 & 0.52 & 0.82 & 0.62 & 7.79 & 6.13 \\
            & 14b & 0.60 & 0.55 & 0.85 & 0.76 & 8.69 & 6.58 \\
            & 32b & \textcolor{red}{0.64} & \textcolor{red}{0.57} & \textcolor{red}{0.91} & \textcolor{red}{0.83} & \textcolor{red}{9.13} & 7.23 \\
            \midrule
            \multirow{3}{*}{Llama 2~\cite{touvron2023llama}} & 7b & 0.52 & 0.48 & 0.81 & 0.62 & 8.02 & 6.10 \\
            & 13b & 0.55 & 0.51 & 0.83 & 0.73 & 8.46 & 6.53 \\
            & 34b & 0.58 & 0.54 & 0.90 & 0.79 & 8.90 & 7.20 \\
            \midrule
            \multirow{3}{*}{Llama 2 Code~\cite{rozière2023code}} & 7b & 0.55 & 0.52 & 0.81 & 0.64 & 7.96 & 6.16 \\
            & 13b & 0.59 & 0.54 & 0.86 & 0.76 & 8.45 & 6.67 \\
            & 34b & 0.63 & 0.56 & 0.91 & 0.81 & 8.95 & \textcolor{red}{7.30} \\
            \bottomrule
        \end{tabular}
    }
    \caption{Performance comparison of different models on our proposed dataset.}
    \label{tab:model_performance}
\end{table*}

\begin{table*}[t]
    \centering
    \small
    \scalebox{0.85}{  % adjust table 90%
    \begin{tabular}{lccccccccc}
        \toprule
         &                   &           &            & \multicolumn{2}{c}{EM} & \multicolumn{2}{c}{EX} & \multicolumn{2}{c}{LLM-based score} \\
         \cmidrule(r){5-6} \cmidrule(r){7-8} \cmidrule(r){9-10}
Model        & M.K & Q.T.M & C.C & Level I   & Level II   & Level I   & Level II   & Level I          & Level II         \\
        \midrule
\multirow{4}{*}{Qwen2.5 14b ~\cite{yang2024qwen}} & $\times$     & $\times$      & $\times$          & 0.50      & 0.40       & 0.72       &    0.61        & 7.63                 &6.10     \\
         & $\checkmark$   & $\times$           & $\times$          & 0.54      & 0.50       & 0.77      & 0.66       & 7.87              & 6.21              \\
         & $\checkmark$   & $\checkmark$       & $\times$          & 0.58      & 0.53       & 0.82      & 0.70       & 8.32             & 6.37             \\
         & $\checkmark$   & $\checkmark$       & $\checkmark$        & 0.60      & 0.55       & 0.85      & 0.76       & 8.69             & 6.58   \\         
    \bottomrule
    \end{tabular}
    }
    \caption{Ablation study of different modules.}
    \label{tab:ablation_performance}
\end{table*}

\subsection{Evaluation}

\textbf{Exact-Match Accuracy (EM)}~\cite{yu2018spider}. This metric  measures whether all SQL components $\mathbb{C}=\left\{C_k\right\}$ of the predicted SQL query match the ground-truth SQL query. It can be computed as follows:

\vspace{-5mm}
\begin{equation}
    E M=\frac{\sum_{i=1}^N \mathbb{I}\left(\bigwedge_{C_k \in \mathbb{C}} Y_i^{C_k}=\hat{Y}_i^{C_k}\right)}{N}
\end{equation}

\noindent \textbf{Execution Accuracy (EX)}~\cite{yu2018spider}. This metric evaluates the performance by comparing whether the execution result sets of the ground-truth and predicted SQL queries are identical. It can be computed as:

\vspace{-5mm}
\begin{equation}
    EX = \frac{\sum_{i=1}^N \mathbb{I}\left(V_i = \hat{V}_i\right)}{N}
\end{equation}

where $\mathbb{I}(\cdot)$ is an indicator function that equals 1 if the condition inside is satisfied, and 0 otherwise. 
%Note that false negatives may occur when two semantically different SQL queries produce identical execution result sets, which could lead to a mistaken assumption that the model’s output is correct.

\noindent \textbf{LLM-based Score} For long-text answers, such as listing medications or responding to hospitalization reports, the previous metrics are not suitable. Inspired by works LLaVa-Med~\cite{li2023llavamed, kweon2024ehrnoteqa}, we use GPT-4~\cite{achiam2023gpt} to evaluate the accuracy of model-generated answers. The reference answer is manually created and serves as the upper bound. GPT-4 then evaluates the model’s output by comparing it to the reference answer. It then assigns a score on a scale from 1 to 10, where 1 indicates poor accuracy and 10 reflects a highly accurate response.

%%%
\begin{comment}
\begin{algorithm}
\caption{Algorithm Framework}
\label{alg:medical_process}
\begin{algorithmic}
    \State \textbf{Input:} EHR databse $\mathcal{D}$, Input question $q \in \mathcal{Q}$, Column description of EHR $\mathcal{D}$: $\mathcal{C}$, Tool set $\mathcal{T}$, Question samples $\mathcal{Q}_{s}$, Medical knowledge $\mathcal{M}$, Generation prompt $\mathcal{P}$.
    \State {We have guided prompt $\mathcal{I} = [\mathcal{C}, \mathcal{M}, \mathcal{P}]$.}
    \State Initialize: $\text{try\_time}: k \in \{1, \dots, K\}$, $\text{flag} = 0$.
    \State {\textcolor{blue}{\% Match similar question examples}}
    \State $q_{sim} = \arg \text{TopK}_{max}(sim(q, q_i | q_i \in \mathcal{Q}_{s})$
    \State {\textcolor{blue}{\% Generate Query Code}}
    \State $S(q) = \text{LLM}(\mathcal{I}, \mathcal{T}, q, q_{sim})$
    \State {\textcolor{blue}{\% Loop until max iterations or successful execution}}
    \While{$k \leq K$ \textbf{and} $\text{flag} = 0$}
        \State {\textcolor{blue}{\% Code Execution}}
        \State $O(q) = \text{EXECUTOR}(S(q))$
        \State {\textcolor{blue}{\% Code Check}}
        \If{$O(q)$ includes error information}
            \State $S(q) = \text{LLM}(S(q), \mathcal{I}, \mathcal{T}, q, \text{error\_info})$
            \State $k = k + 1$
        \Else
            \State $\text{flag} = 1$
        \EndIf
    \EndWhile
    \State \textbf{Output:} Final answer or output information from $O(q)$ 
\end{algorithmic}
\end{algorithm}
\end{comment}
%%%

\section{Experiment}
\label{sec:experiment}

%%%
\begin{comment}
\begin{table*}
  \centering
  \begin{tabular}{lll}
    \hline
    \textbf{Method} & \textbf{MIMIC-IV} & \textbf{eICU} \\
    \hline
    Method 1        & 10           &                           \\
    Method 2        & 10        &                           \\
    Method 3        & 10     &                           \\
    Method 4        & 10     &                           \\
    Method 5        & 10      & 10          \\
    \hline
  \end{tabular}
  \caption{\label{citation-guide}
    Main results of success rate on different datasets.
  }
\end{table*}
\end{comment}
%%%

\subsection{Experiment Setup}

\textbf{Task and Datasets.} We use test data from our constructed dataset, which includes 1000 Level I and 1000 Level II questions (see Appendix~\ref{tab:dataset split}). The task is to evaluate the accuracy of the generated query statements and the correctness of the query results. Questions are categorized into two levels based on difficulty: Level I for those with no more than three constraint conditions, and Level II for those with more than three.

\noindent \textbf{Model Select.} %{\color{red}\sout{We used different large models as query generation models, including the Qwen2.5}}~\cite{yang2024qwen} {\color{red}\sout{series and the LLaMA}}~\cite{touvron2023llama} {\color{red}\sout{series. We also used models with different parameter amounts to measure their impact on query generation capabilities. These models are listed below:}}
Under our proposed TQGen-EHRQuery framework, we utilized various large models for query generation, including the Qwen2.5~\cite{yang2024qwen} and LLaMA~\cite{touvron2023llama} series, with different parameter sizes to assess their impact on query generation. The models used are:

\begin{enumerate}[itemsep=0pt, parsep=0pt, topsep=0pt]
    \item Qwen2.5-7B/14B/32B~\cite{yang2024qwen} is for general-purpose language understanding and generation tasks.
    \item Qwen2.5 Code-7B/14B/32B~\cite{hui2024qwen2} is an optimized version of Qwen 2.5 tailored specifically for programming-related tasks.
    \item LLaMA 2-7B/13B/34B~\cite{touvron2023llama} is for general text understanding task.
    \item LLaMA 2 Code 7B/13B/34B~\cite{rozière2023code} is a variant of Llama 2 fine-tuned for coding tasks.
\end{enumerate}

\noindent \textbf{Implementation Details.} The experiments were conducted on an NVIDIA GeForce RTX A6000 GPU. To ensure consistency, we set the temperature parameter to 0 during API calls to GPT-4, eliminating randomness in the generated responses. The generated queries are in SQL format, and their execution is facilitated using Python.

\begin{figure*}[htbp]  % htbp 控制图片浮动的位置
  \centering  % 图片居中
  \includegraphics[width=0.95\textwidth]{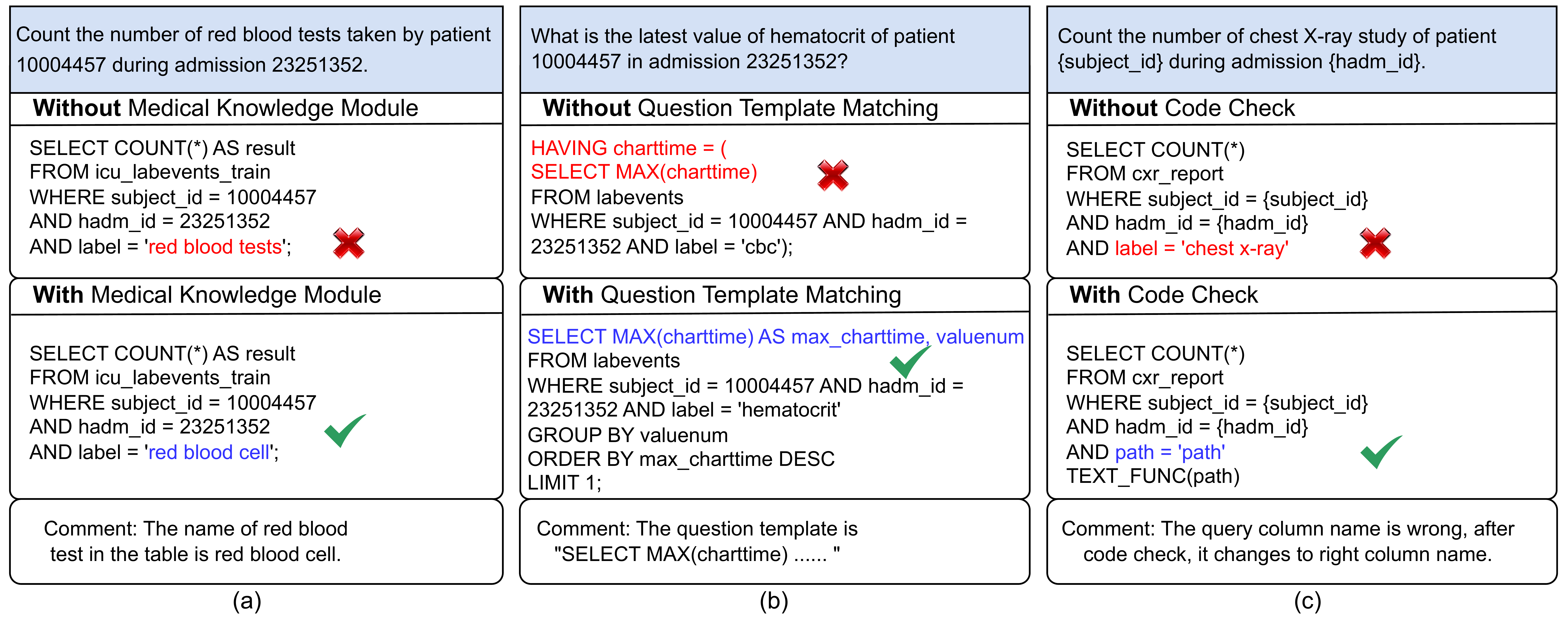}  % 调整图片大小并插入图片
  \caption{Some examples demonstrate the efficiency of the modules. The top row shows the questions, followed by the generated query codes in the second and third rows — one without the module and the other with the module. The last row explains why the query code is correct.}  % 图片标题
  \label{fig:result example}  % 用于引用图片
\end{figure*}

\subsection{Quantitative Analysis}

We evaluated the performance of various models on the dataset using three metrics: exact match accuracy (EM), execution accuracy (EX), and a large language model-based score (LLM-based score). Exact match accuracy and execution accuracy were employed to assess the correctness of results involving simple data types, such as numerical values and strings. In contrast, the LLM-based score was specifically designed to evaluate tasks that involve complex text comprehension and generation. 

Table~\ref{tab:model_performance} summarizes the experimental results. As shown in the table, with an increase in the model’s parameter count, the model’s performance on EX, EX, and LLM-based scores improves. Additionally, when the questions are relatively simple, the accuracy of the generated query code and its execution results is higher. However, as the difficulty of the questions increases, the decline in EX for the generated query code becomes more significant, while the decrease in EX is less pronounced. This is because complex problems require more function combinations, and although the model uses different function combinations, it ultimately achieves the same result. For text-based query tasks, the LLM-based score also experiences a decline, likely due to the complexity of the questions causing the model to select incorrect texts, thereby affecting accuracy.

%{\color{red}\sout{We also investigated the differential impact of various functional modules on the overall performance of the dataset. In previous sections, we introduced several key components, including table description, medical knowledge, question template matching, and code checking. For the experimental design, we configured the agent with table descriptions and prompts, and subsequently evaluated the specific influence of the three modules—medical knowledge (M.K), question template matching (Q.T.M), and code checking (C.C)—on the generated query code.}}

We also investigated the impact of various functional modules within the TQGen-EHRQuery framework on the overall performance of the dataset. For the experimental design, we configured the agent with table descriptions and prompts, then evaluated the specific effects of three modules—Medical Knowledge (M.K), Question Template Matching (Q.T.M), and Code Inspection (C.C)—on the generated query code.

We employed the Qwen2.5-14B~\cite{yang2024qwen} model as the foundation and randomly sampled 500 instances from a self-constructed test set, encompassing samples with two distinct levels of difficulty. Comparative analysis revealed a significant decrease in the model’s accuracy when the M.K, Q.T.M, and C.C modules were disabled.

Disabling the medical knowledge module led to the most significant accuracy decline, likely due to mismatches between query phrases and table entries, causing retrieval failures. The question template matching module also had a notable impact on code generation accuracy, with matched questions and exemplar code boosting performance. However, the model occasionally produced correct code without Q.T.M support. In contrast, the code checking module had a smaller effect, as most SQL queries executed correctly without modification, with adjustments needed only in specific edge cases.

\subsection{Case Study}

The effectiveness of the aforementioned modules is demonstrated through the experimental results. As illustrated in Fig~\ref{fig:result example}, the impact of different modules on code generation and query result generation is depicted across three subplots: (a) shows the performance differences with and without the medical knowledge (M.K) module, indicating a significant improvement in the accuracy of medical term matching when the module is enabled; (b) compares the outcomes with and without the question template matching (Q.T.M) module, highlighting the crucial role of template matching in the task; and (c) validates the contribution of the code checking (C.C) module. The experimental results confirm that all three modules contribute to enhanced accuracy and reliability of the query results.

\section{Conclusion}
\label{sec:conclusion}

In this work, we present a novel approach for querying EHRs by integrating structured tables and unstructured clinical text. We created a publicly available dataset to facilitate natural language-to-query translation, addressing the complexities of EHR data, including complex table relationships, long-form narratives, and specialized medical terminology. Additionally, we propose a workflow leveraging LLMs, incorporating modules for medical knowledge, question templates, and toolsets to enhance query accuracy. Our findings demonstrate that LLM-powered querying systems can significantly improve EHR data accessibility and usability, paving the way for more efficient clinical information retrieval. Future work will focus on enhancing query accuracy, incorporating multi-modal data sources, and further validating the approach in real-world clinical settings.

\section*{Limitation}
\label{sec:limitation}

%{\color{red}\sout{Despite careful design, our dataset has some limitations. Since it is based on the MIMIC database, its generalizability may be restricted, which could affect the stability, comprehensiveness, and applicability of our model. In addition, although the current query module can handle medical-related terms well and improve query results, there is still much room for improvement in query performance in complex situations. Future work should address these challenges. While our research represents a significant step in multimodal EHR QA systems, there is still room for improvement. Key future directions include expanding the dataset by enhancing multimodal dialogue systems and integrating mechanisms to handle unanswerable or ambiguous questions, which are crucial for real-world applications. These efforts will leverage our dataset as a valuable resource and lay the foundation for more comprehensive healthcare solutions.}}

Despite careful design, our dataset has some limitations due to its reliance on the MIMIC database, which may limit its generalizability and impact the model’s stability, comprehensiveness, and applicability. Additionally, while our current query module effectively handles medical terminology and structured queries, it still faces challenges in complex problem scenarios, particularly when dealing with rare conditions, complex limit condition, or ambiguous queries that require contextual understanding beyond the structured data. Future work will focus on expanding the dataset, enhancing multimodal dialogue systems, and developing mechanisms to address unanswerable or ambiguous questions—critical for real-world applications. These efforts will leverage our dataset as a valuable resource and lay the foundation for more comprehensive healthcare solutions.

\section*{Ethical and Privacy Considerations}

In accordance with the PhysioNet Certified Health Data Use Agreement, we strictly prohibit transferring confidential patient data (MIMIC-IV) to third parties, including via online services like APIs. To ensure compliance, we use locally deployed models for testing, preventing third-party access to sensitive patient information. We continuously monitor our adherence to these guidelines and relevant privacy laws to ensure ethical data use. Sensitive information, such as patient names and visit times, has been appropriately processed to protect patient privacy.

\section*{Acknowledgments}

We thank the MIMIC-IV and MIMIC-IV-Note datasets for providing valuable clinical and textual data that support our research. These resources have been essential in developing and evaluating our multi-modal EHR QA system. 
%We also appreciate the MIT Laboratory for Computational Physiology for curating and maintaining these datasets, driving advancements in clinical NLP and healthcare AI.

\bibliography{custom}

\appendix

%\clearpage
\section{Appendix}
\label{sec:appendix}

\subsection{EHR Dataset Introduction}
\label{sec:ehr dataset introduction}

The \textbf{MIMIC-IV (v2.2)} dataset ~\cite{johnson2023mimiciv} is a large, publicly accessible relational database containing de-identified health-related data, including diagnoses, procedures, and treatments, for 50,920 patients who were admitted to the critical care units of Beth Israel Deaconess Medical Center (BIDMC) between 2008 and 2019.

\noindent The \textbf{MIMIC-CXR} dataset ~\cite{johnson2019mimiccxr} is a large-scale, publicly available collection of 377,110 chest radiographs from 227,827 imaging studies conducted at BIDMC between 2011 and 2016. MIMIC-CXR can be linked to MIMIC-IV through lookup tables that map patient identifiers across the two datasets.

\noindent The \textbf{MIMIC-IV-Note} dataset ~\cite{johnson2023mimiciv} is a de-identified collection of free-text clinical notes linked to the MIMIC-IV database. It comprises 331,794 discharge summaries from 145,915 patients (both hospital and emergency department admissions) and 2,321,355 radiology reports from 237,427 patients. All notes have been de-identified in accordance with HIPAA Safe Harbor standards.

We also list the tables and columns used in our dataset in Table~\ref{tab:tabels and columns}. There are 18 tables, and all tables are linked by the subject\_id and hadm\_id. In the MIMIC-CXR dataset, the path of radiology report are stored in the path column, and the discharge summary are stored in the text column in the MIMIMC-Note.

% \usepackage{multirow}
\begin{table*}[h]
    \small
    \begin{tabular}{|c|c|l|p{8cm}|}
    \hline
    \textbf{Index} & \textbf{Dataset}                    & \textbf{Table}              & \textbf{Columns}  \\
    \hline
    1     & \multirow{15}{*}{MIMIC-IV} & patients           & subject\_id, hadm\_id, gender, anchor\_age, anchor\_year, dod. \\
    2     &                            & admissions         & subject\_id, hadm\_id, admittime, dischtime, admission\_type, admission\_location, discharge\_location, insurance, marital\_status, race. \\
    3     &                            & diagnoses          & subject\_id, hadm\_id, icd\_code, icd\_version. \\
    4     &                            & d\_icd\_diagnoses  & icd\_code, icd\_version, long\_title.  \\
    5     &                            & labevents          & subject\_id, hadm\_id, item\_id, charttime, valuenum, valueuom, ref\_range\_lower, ref\_range\_upper.                                     \\
    6     &                            & d\_labitems        & itemid, label, fluid, category.  \\
    7     &                            & microbiolog & subject\_id, hadm\_id, charttime, spec\_type\_desc, test\_name. \\
    8     &                            & prescriptions      & subject\_id, hadm\_id, starttime, stoptime, drug, dose\_val\_rx, dose\_unit\_rx, route.      \\
    9     &                            & procedures         & subject\_id, hadm\_id, icd\_code, icd\_version. \\
    10    &                            & d\_icd\_procedures & icd\_code, icd\_version, long\_title    \\
    11    &                            & icustays           & subject\_id, hadm\_id, stay\_id, first\_careunit, last\_careunit, intime, outtime, los.      \\
    12    &                            & inputevents        & subject\_id, hadm\_id, stay\_id, starttime, itemid, amount, amountuom,patientweight, etc.      \\
    13    &                            & d\_items           & itemid, label, abbreviation, category, unitname. \\
    14    &                            & outputevents       & subject\_id, hadm\_id, stay\_id, charttime, itemid, value, valueuom.  \\
    15    &                            & chartevents        & subject\_id, hadm\_id, stay\_id, charttime, itemid, value, valueuom. \\
    \hline
    16    & \multirow{2}{*}{MIMIC-CXR} & cxr-metadata & subject\_id, tudy\_id, dicom\_id, studydate, studytime. \\
    17    &                            & cxr-record-list    & subject\_id, study\_id, dicom\_id, path.   \\
    \hline
    18    & MIMIC-IV-Note              & discharge          & subject\_id, hadm\_id, charttime, storetime, text.    \\ 
    \hline                                   
    \end{tabular}
    \caption{Dataset, tables, and columns used in our dataset construction.}
    \label{tab:tabels and columns}
\end{table*}

\subsection{Dataset Construction}
\label{sec:appendix dataset construction}

%{\color{red}\sout{During the dataset construction process, we employed a dual-faceted question design strategy. First, we formulate question templates based on consultations with clinical experts and insights from the relevant literature. Second, we specifically designed complex questions that require the integration of structured data (e.g., tabular information) with unstructured textual data (e.g., radiology reports and discharge summaries). For each standardized question template, we systematically generated multiple paraphrased variants that maintain semantic equivalence, thereby enhancing the diversity and comprehensiveness of the question set. Finally, a sampling approach was used to randomly select one variant from the pool of candidate questions, which was then populated with relevant field values to generate the final question presented to the research subjects.}} 

We first constructed a series of question templates related to EHR data, covering patient information, laboratory indicators, medication usage, length of hospitalization, discharge details, and radiology reports. Based on physicians’ advice, we refined and optimized these templates to ensure their clinical relevance. Subsequently, for each template, we used ChatGPT to generate multiple rephrasings, which were then manually reviewed and filtered, ultimately retaining approximately ten distinct variations per template to enhance question diversity. For table-based questions, we referred to the question types in EHRSQL and, following physicians’ suggestions, incorporated inquiries regarding laboratory indicator values, trends, statistical measures, and whether they fell within normal ranges. These questions were considered highly valuable in clinical practice. For text-based questions, we check the types of information contained in the text, such as patient history, admission reasons, discharge prescriptions, and discharge status in discharge summaries. Based on this analysis, we designed corresponding question templates and ensured their clinical relevance under physicians’ guidance.

For data available in both tables and discharge reports, such as blood test results, the discharge reports often lack precise timestamps, as illustrated in Textbox~\ref{box:discharge_summary_blood}. To address this, we designed prompts that guide the model to prioritize retrieving answers from structured table data. For text-based questions, we explicitly instructed the model to extract answers from textual data within the question templates, ensuring consistency and minimizing potential biases arising from discrepancies between data sources. Table~\ref{tab:question examples ehr tables} lists some question examples related to EHR tables. Table~\ref{tab:question examples cxr report} lists some questions related to the text of the CXR reports. Table~\ref{tab:question examples discharge summary} lists some question examples related to the text of the discharge summaries.

The question answer example is listed in ~\ref{box:question answer example}. "question\_template" is the template question, "question" is the real question filled with values. "query\_code" is the generated query code. "answer" is the answer after running the query code.

\begin{table*}[]
    \small
    \begin{tabular}{|c|p{13cm}|}
    \hline
    \textbf{Related Table} & \textbf{Question Template} \\ \hline
    diagnoses & Has patient \{subject\_id\} been   diagnosed with \{diagnoses\_name\} during admission \{hadm\_id\}?                       \\
    \hline
    admission & List the hospital admission time of patient \{subject\_id\}.                                                               \\
    \hline
    icu\_stay & Count the number of ICU visits of patient \{subject\_id\} during admission \{hadm\_id\}.                                 \\
    \hline
    labevents & Count the number of \{labtest\_name\} patient \{subject\_id\} received during admission \{hadm\_id\}.                    \\
    \hline
    labevents & For patient \{subject\_id\} in admission \{hadm\_id\}, what was the highest value of \{labtest\_name\}?                  \\
    \hline
    microbiolog & What are the top {[}n\_rank{]} frequent microbiology tests that patient \{subject\_id\} had  in admission \{hadm\_id\}?  \\
    \hline
    prescriptions & What are the top {[}n\_rank{]} frequently prescribed drugs of  \{gender\_type\} patients aged \{age\_group\} in \{year\}? \\
    \hline
    labevents & For patient \{subject\_id\} in admission \{hadm\_id\}, was the last \{labtest\_name\} normal?                            \\
    \hline
    prescriptions & Has patient \{subject\_id\} have \{durg\_name\} in his/her \{ordinal\_num\} admission?       \\
    \hline
    microbiology & What was the time that patient \{subject\_id\} have \{microbiology\_name\} in his/her \{ordinal\_num\} admission?       \\
    \hline
    \end{tabular}
    \caption{Question templates examples related to EHR tables.}
    \label{tab:question examples ehr tables}
\end{table*}

\begin{table*}[]
    \small
    \begin{tabular}{|c|p{13cm}|}
    \hline
    \textbf{Related Table} & \textbf{Question Template} \\ \hline
    cxr\_report & List the \{findings\_name\} of the last chest X-ray study for patient \{subject\_id\} during the hospital stay within admission \{hadm\_id\}. \\
    \hline
    cxr\_report & List the \{study\_date\} of patient \{subject\_id\} who had a chest X-ray study during hospital visit indicating \{findings\_name\} within the admission \{hadm\_id\}.\\
    \hline
    cxr\_report & List the \{findings\_name\} of the chest X-ray study \{study\_id\} for patient \{subject\_id\} during the admission \{hadm\_id\}. \\
    \hline
    cxr\_report & Count the number of chest X-ray study of patient \{subject\_id\} during admission \{hadm\_id\}. \\
    \hline
    cxr\_report & Count the number of \{gender\} patients aged \{age\_group\} who had a chest X-ray study during hospital visit indicating \{findings\_name\} in the \{year\}.  \\
    \hline
    cxr\_report & Has patient \{subject\_id\} been diagnosed with \{diagnoses\_name\} and also had a chest X-ray study indicating \{findings\_name\} within the admission \{hadm\_id\}?\\
    \hline
    cxr\_report & Has patient \{subject\_id\} received a \{procedure\_name\} procedure and also had a chest X-ray study\ indicating \{findings\_name\} in the \$\{natomical\_area\} within the admission \{hadm\_id\}? \\
    \hline
    cxr\_report & Has patient \{subject\_id\} been prescribed with \{drug\_name\} and also had a chest X-ray study indicating \{findings\_name\} in the \$\{anatomical\_area\} within the admission \{hadm\_id\}? \\ 
    \hline
    \end{tabular}
    \caption{Question templates examples related to CXR report text.}
    \label{tab:question examples cxr report}
\end{table*}

\begin{table*}[]
    \small
    \begin{tabular}{|c|p{13cm}|}
    \hline
    \textbf{Related Table} & \textbf{Question Template} \\ \hline
    discharge       & According to the \{ordinal\_num\} discharge summary of patient   \{subject\_id\}, does the patient \{subject\_id\} have any known drug allergies?    \\ \hline
    discharge       & According to the \{ordinal\_num\} discharge summary of patient   \{subject\_id\}, what was the patient \{subject\_id\} primary reason for admission?    \\ \hline
    discharge       & According to the \{ordinal\_num\} discharge summary of patient   \{subject\_id\}, what was the patient \{subject\_id\} discharge diagnosis?         \\ \hline
    discharge       & According to the \{ordinal\_num\} discharge summary of patient   \{subject\_id\}, what medications were prescribed to the patient \{subject\_id\}   upon discharge?   \\ \hline
    discharge       & According to the \{ordinal\_num\} discharge summary of patient   \{subject\_id\}, what is the family history of the patient \{subject\_id\}?                          \\ \hline
    discharge       & According to the \{ordinal\_num\} discharge summary of patient   \{subject\_id\}, describe the hospital course briefly.                                               \\ \hline
    discharge       & According to the \{ordinal\_num\} discharge summary of patient   \{subject\_id\}, what medication on admission is given to the patient?                               \\ \hline
    discharge       & According to the \{ordinal\_num\} discharge summary of patient   \{subject\_id\}, what was the discharge disposition of the patient?                                 \\ \hline
    discharge       & According to the \{ordinal\_num\} discharge summary of patient   \{subject\_id\}, was the patient's condition improving?                                              \\ \hline
    discharge       & According to the \{ordinal\_num\} discharge summary of patient   \{subject\_id\}, list all the blood test items the patient have taken.                               \\ \hline
    discharge       & According to the \{ordinal\_num\} discharge summary of patient   \{subject\_id\}, what happened to the labtest \{blood\_test\_item\}?                                 \\ \hline
    discharge       & According to the \{ordinal\_num\} discharge summary of patient   \{subject\_id\}, why the \{blood\_test\_item\} change?                                               \\ \hline
    discharge       & According to the \{ordinal\_num\} discharge summary of patient   \{subject\_id\}, did the patient receive labtest \{blood\_test\_item\}?                             \\ \hline
    \end{tabular}
    \caption{Question templates examples related to discharge summary text.}
    \label{tab:question examples discharge summary}
\end{table*}

\begin{tcolorbox}[colback=gray!5!white, colframe=gray!75!black, title={Incomplete blood test information in discharge summary}]
    \label{box:discharge_summary_blood} % Placed right after title
    \small
    Pertinent Results: \\
    \textbf{ADMISSION LABS:} \\
    \_\_\_ 02:40AM BLOOD WBC-15.0* RBC-3.75* Hgb-11.6 Hct-36.1 \\
    MCV-96 MCH-30.9 MCHC-32.1 RDW-13.7 RDWSD-48.0* Plt \_\_\_ \\
    \_\_\_ 02:40AM BLOOD Neuts-87.9* Lymphs-5.3* Monos-5.8  \\
    Eos-0.0* Baso-0.1 Im \_\_\_ AbsNeut-13.14* AbsLymp-0.80*  \\
    AbsMono-0.87* AbsEos-0.00* AbsBaso-0.01  \\
    \_\_\_ 02:40AM BLOOD Glucose-128* UreaN-12 Creat-0.4 Na-137  \\
    K-3.8 Cl-100 HCO3-25 AnGap-12  \\
    \_\_\_ 02:40AM BLOOD cTropnT-<0.01  \\
    \_\_\_ 05:40AM BLOOD Calcium-8.9 Phos-3.2 Mg-2.2  \\
    \_\_\_ 02:52AM BLOOD Glucose-128* Creat-0.4 Na-136 K-3.6  \\
    Cl-103 calHCO3-24  \\
    \label{box:discharge summary blood}
\end{tcolorbox}

\begin{tcolorbox}[colback=gray!5!white, colframe=gray!75!black, title=Prompt,]
    \small
    Please generate python code to answer the question. \\
    Use pandas package to load .csv or .csv.gz file. \\
    Only generate code for the question. \\
    No explanation and other description. \\
    Use `print` to output the result.  \\
    The final result variable should be named as `result`. \\
    Questions related to discharge summary should be answered based on the summarytext. \\
    \label{box:prompt example}
\end{tcolorbox}

\subsection{Dataset Statistics}

Here we list some statistics information for our constructed dataset in~\ref{tab:dataset split}. We classify the difficulty level of the questions based on the number of values that need to be filled in when querying. Questions that require no more than three fill-in values are classified as Level 1 questions, and questions that require more than three fill-in values are classified as Level 2 questions. The more fill-in values required, the more complex the question.

\begin{table*}
  \centering
  \small
    \begin{tabular}{|l|ll|ll|ll|}
    \hline
                      & \multicolumn{2}{c|}{Train}           & \multicolumn{2}{c|}{Valid}           & \multicolumn{2}{c|}{Test}            \\ \hline
                      & \multicolumn{1}{l|}{Level I} & Level II & \multicolumn{1}{l|}{Level I} & Level II & \multicolumn{1}{l|}{Level I} & Level II \\ \hline
    Table             & \multicolumn{1}{l|}{2000}   & 2000   & \multicolumn{1}{l|}{500}    & 500    & \multicolumn{1}{l|}{500}    & 500    \\ \hline
    CXR report        & \multicolumn{1}{l|}{1000}   & 1000   & \multicolumn{1}{l|}{250}    & 250    & \multicolumn{1}{l|}{250}    & 250    \\ \hline
    Discharge         & \multicolumn{1}{l|}{1000}   & 1000   & \multicolumn{1}{l|}{250}    & 250    & \multicolumn{1}{l|}{250}    & 250    \\ \hline
    Total             & \multicolumn{1}{l|}{4000}   & 4000   & \multicolumn{1}{l|}{1000}   & 1000   & \multicolumn{1}{l|}{1000}   & 1000   \\ \hline
    \end{tabular}
    \caption{Statistics of our dataset}
    \label{tab:dataset split}
\end{table*}

\subsection{Prompt Detail}
\label{sec:prompt detail}

For each table, we provide a detailed explanation of the information conveyed by the table and specify the exact file path from which the table can be accessed. Additionally, for each column within the table, we offer a comprehensive definition that clarifies the specific meaning and significance of the data it represents. Textbox~\ref{box:table description} gives an example of the description to admissions table.

When the model encounters a question it cannot answer, we introduce a mechanism in the prompt design to ensure a predefined response. Specifically, the model is instructed to return a default value, such as “No corresponding information found” when it cannot generate a valid query, see Textbox~\ref{box:question cannot answer}. This fallback approach improves robustness by providing consistent feedback, even when the question does not match the database schema. By integrating this method, the system remains reliable and predictable, particularly for edge cases or unanswerable queries.

\begin{tcolorbox}[colback=gray!5!white, colframe=gray!75!black, title=Question cannot answer,]
    \small
    \# Background. This patient didn't have RBC test. \\
    Question: What is the RBC value of patient 10054388 in the first admission? \\
    Answer: No corresponding information found. \\
    \label{box:question cannot answer}
\end{tcolorbox}

\begin{tcolorbox}[colback=gray!5!white, colframe=gray!75!black, title=Table Description,]
    \small
    This is the description to the admissions.csv.gz file. This file is located in mimic-iv/admissions.csv.gz. \\
    \textbf{subject\_id}: A unique identifier for each patient in the dataset. Each patient only has one subject\_id. \\
    \textbf{hadm\_id}: Hospital admission ID, a unique identifier for each hospital admission. This ID enables differentiation between multiple admissions for the same patient. \\
    \textbf{admittime}: Timestamp for the exact date and time when the patient was admitted to the hospital. This helps establish the start of a hospital stay. \\
    \textbf{dischtime}: Timestamp for the date and time when the patient was discharged from the hospital, marking the end of a specific admission period. \\
    \textbf{admission\_type}: Categorical field indicating the type of admission, such as "emergency," "urgent," or "elective." This provides context on the reason or urgency of admission. \\
    \textbf{admission\_location}: Describes the location from which the patient was admitted, such as "clinic referral," "emergency department," or "transfer from another facility." \\
    ...... \\
    \label{box:table description}
\end{tcolorbox}

\begin{tcolorbox}[colback=gray!5!white, colframe=gray!75!black, title=Question Answer Example,]
    \small
    "subject\_id": 10054277, \\
    "hadm\_id": 27607912, \\
    "question\_answer\_pairs": [ \\
        \{"question\_template": "Count the admission num of patient \{subject\_id\}.", \\
        "question": "How many times does the record show regarding patient 10054277's admissions?", \\
        "query\_code":  df = pd.read\_csv ('patients.csv.gz') \\
                        result = df[df['subject\_id'] == 10054277] ['hadm\_id'].nunique() \\
                        print(result) \\
        "answer": "1" \} \\
        \\
    "subject\_id": 10054277, \\
    "hadm\_id": 27607912, \\
    "question\_answer\_pairs": [ \\
        \{"question\_template": "What diseases does the patient subject\_id have in the admission admission\_id according to the radiology report?", \\
        "question": "According to the radiology report, what diseases are associated with patient 10054277 in admission 27607912?", \\
        "query\_code":  df = pd.read\_csv ('patients.csv.gz') \\
                        patient\_data = df[(df['subject\_id'] == 10054277) \& (df['hadm\_id'] == 27607912)] \\
                        report\_path = patient\_data["path][0] \\
                        question\_text = "What disease in the CXR report?" \\
                        result = text\_func(report\_path, question\_text)
        "answer": "atelectasis, pleural effusion." \} \\
    \label{box:question answer example}
\end{tcolorbox}

\end{document}